\documentclass[aps,prb,showpacs,superscriptaddress,twocolumn]{revtex4}

\usepackage{graphicx}
\usepackage{amssymb,amsmath}
\usepackage{verbatim}

\def\Mvariable#1{#1}

\begin{document}
\newcommand{\sgn}{\,{\rm sgn}\,}
\setlength{\unitlength}{1cm}

\title{Transport of fractional Hall quasiparticles through an antidot}

\author{Matteo Merlo}
\affiliation{Dipartimento di Fisica, INFN, Universit\`{a} di Genova,
   Via Dodecaneso 33, I-16146 Genova, Italy}
\affiliation{Dipartimento di Fisica, LAMIA-INFM-CNR, Universit\`{a} di Genova,
   Via Dodecaneso 33, I-16146 Genova, Italy}
\author{Alessandro Braggio}
\affiliation{Dipartimento di Fisica, LAMIA-INFM-CNR, Universit\`{a} di Genova,
   Via Dodecaneso 33, I-16146 Genova, Italy}
\author{Nicodemo Magnoli}
\affiliation{Dipartimento di Fisica, INFN, Universit\`{a} di Genova,
   Via Dodecaneso 33, I-16146 Genova, Italy}
\author{Maura Sassetti}
\affiliation{Dipartimento di Fisica, LAMIA-INFM-CNR, Universit\`{a} di Genova,
   Via Dodecaneso 33, I-16146 Genova, Italy}

\date{\today}

\begin{abstract}
Current statistics of
an antidot in the  fractional quantum Hall regime is studied for Laughlin's series.
The chiral Luttinger liquid picture of edge states with a renormalized interaction exponent $g$ is
adopted. Several peculiar features are found in the sequential tunneling regime. On one side, current displays
negative differential conductance and double-peak structures when $g<1$.
On the other side, universal sub-poissonian transport regimes are identified
through an analysis of higher current moments.
A comparison between Fano factor  and skewness is proposed in order to clearly distinguish the charge of the carriers,
regardless of possible non-universal interaction renormalizations.
Super-poissonian statistics is obtained in the shot limit for $g<1$, and
plasmonic effects due to the finite-size antidot are tracked.
\end{abstract}


\maketitle

\section{Introduction}
The peculiar properties of quasiparticles (qp) in the fractional quantum
Hall effect (FQHE) have received great attention
especially for the states at filling factor $\nu=1/p, \, p$ odd integer,
in which gapped bulk excitations were predicted
to exist and to possess fractional charge\cite{Laughlin} $e^*=\nu e$
($e<0$ electron charge) and statistics.\cite{HalperinArovas}

A boundary restriction of this theory was subsequently put forward
in terms of edge states by Wen.\cite{Wen} This theory recovered the
fractional numbers of quasiparticles in the framework of chiral
Luttinger Liquids ($\chi$LL), and indicated tunneling as an
accessible tool to probe them.\cite{kanefisher}
Accordingly, quasiparticles with charge $e/p$  were measured in shot noise experiments with point contact
geometries and edge-edge backscattering.\cite{depicciotto}

A key prediction of $\chi$LL theories is that the interaction parameter should be universal and equal to $\nu$.
As a consequence, the quasiparticle (electron) local tunneling density of states obeys
a power law in energy $D\propto E^{\nu^{}-1}$ ($D\propto E^{1/\nu^{}-1}$).
Several geometries have been set up in experiments to test this nonlinearity through measurements
of tunneling current $I$ versus bias voltage $V$. For instance, in the case of electron tunneling between a metal
and an edge at filling $\nu$, one should have $I\propto V^{\alpha}$
in the limit $eV\gg k_B T$ with $\alpha=1/\nu$.\cite{kanefisher}
Experiments\cite{chang} at filling factor $1/3$ indeed proved a power-law behaviour but with $\alpha\neq3$.
Deviations were observed also with quasiparticle tunneling in an almost open point contact geometry
 at $\nu=1/3$;\cite{chung} here, the predicted backscattering linear conductance is $G_B \propto T^{2\nu-2}$
while the measured quantity obeys a power law with positive exponent.
Analogous discrepancies were observed in a similar geometry, with the quasiparticle tunneling differential conductance developing a minimum
around zero bias instead of a maximum for decreasing temperature.\cite{beltram}
Moreover, several numerical calculations and simulations also disagree with the conventional chiral Luttinger theories. We mention for instance
finite-size exact-diagonalization calculations with short-range\cite{zuelicke} or  Coulomb electron-electron
interactions\cite{tsiperg,goldmant,cheianov} and Monte Carlo simulations with interacting Composite Fermions on a ring.\cite{mandaljain}

The disagreements of $\chi$LL predictions with observed exponents are still not completely understood,
although several theoretical mechanisms have been put forward to reproduce a renormalized Luttinger parameter,
including coupling to phonons or dissipative environments,\cite{halpros,eggert,khlebnikov,levitov}
effects of interaction range,\cite{papa,evers} edge reconstruction with smooth confining potentials.\cite{yang,wan,joglekar,papa2}

Our purpose is to discuss fractional Hall edges in an enriched $\chi$LL theory where the possibility of a renormalized
interaction parameter $g\neq \nu$ is assumed, analysing different transport regimes
and clearly distinguishing signatures of charge $\nu e$ qp from effects due to the quasiparticle propagators governed by $g$.
To do so, we choose a geometry where two fractional quantum Hall edges are connected via weak tunneling through an edge state encircling an
antidot as in Fig.~\ref{figureone}.\cite{Geller97,merlo}
Such a setup has proven to be extremely versatile and controllable. It has been for instance
employed in a series of experiments where fractional charge and statistics of quasiparticles were
addressed.\cite{gold}
The same geometry has been also used to detect blockade effects
and Kondo physics with spinful edges in the integer regime.\cite{cavendish}

As mentioned before, a series of experimental observations of fractional charge has been based on noise measurements:\cite{depicciotto}
the possibility to extract a significant $e^*/e$ from such measurements
is given by the fact that the geometry of the setup and the tunneling regime ensure a
poissonian process. Smooth evolution in the charge carrier at $\nu=1/3$ from $e^*=\nu e$ to $e$ was
observed in point contacts changing the backscattering amplitude via gates from 0 to 1.\cite{griffiths}
This appears consistent with evolution from fractional qp to electron tunneling.\cite{kanefisher92}
Anyway, this conclusion can only be obtained under the additional hypothesis of  independent particle tunneling.\cite{nota4}
Only the observation of higher moments, as e.g. the normalized skewness, could cross-check this hypothesis.\cite{saleurweiss}
Therefore, to fully explain the charge measurements one has to observe higher moments beyond the Fano factor.
Recent advances in measurement techniques could open this intriguing possibility, especially in view of the unexpected
results recently reported.\cite{chung,beltram}

In this paper we propose to compare Fano factor and skewness in transport regimes where the statistics of the
charge transfer is not poissonian.
We focus our attention on transport through an
antidot in weak-backscattering and sequential tunneling limit, at fractional filling factor $\nu=1/p$ ($p$ odd
integer). Our task is twofold: on one side, we analyse
the tunneling current. It presents remarkable features driven by $g$, like e.g.
negative differential conductance and double-peak structures. Power-law
behaviours exist and can be used to determine the renormalized interaction parameter.
On the other side, we derive a method to assess fractional charge
independently of possibly renormalized $g\neq\nu$.
We analyse noise and skewness in processes with different transport statistics both in the shot and
in the thermal limit. We find universal points that unambiguously define the fractional charge.
In addition, we describe transport regions where the Fano factor is sensitive to the power laws of the quasiparticle propagators and
presents super-poissonian correlations.

The paper is organized as follows: in Sec. II the model is introduced and bosonization procedures
are briefly reviewed. Sections III and IV contain the description of methods and results.
Finally, in Sec. V we discuss our findings and comment existing experimental applications.
\section{model}
\begin{figure}
 \includegraphics[angle=0,width=5. cm]{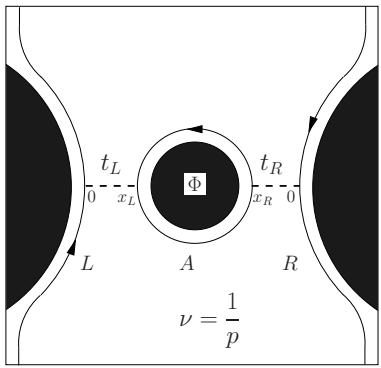}
 \caption{Geometry of the system. In white, the Hall fluid; in
 black, the depleted areas defining the two quasiparticle tunneling points between
 the left ($L$) or right ($R$) edge and the antidot ($A$). In this
 setting, the magnetic field points out of the plane.}
 \label{figureone}
\end{figure}
In our model, edge states form at the boundaries of the sample and around the antidot
(Fig.~\ref{figureone}); mesoscopic effects are associated to the finite size of the antidot
through an Aharonov-Bohm (AB) coupling, and tunneling barriers couple the circular
antidot with both edges. The complete Hamiltonian reads
\begin{equation}\label{bighamil}
H=H^0_{{L}}+H^0_{{R}}+H^0_{{A}}+H^{\rm AB}+H_R^{{T}}+H_L^{{T}},
\end{equation}
and the individual pieces are now described in detail.
\subsection{Bosonization of free Hamiltonians}\label{sec1}
In Eq.~(\ref{bighamil}), $H^0_l$  are standard Wen's hydrodynamical Hamiltonians for the left, right and antidot edge ($l=L,R,A$):
 in terms of the electron excess density $\rho_l(x)$ ($\hbar=1$), one has\cite{Geller97,merlo,kette97}
\begin{equation} \label{hamilt}
H_l^0=\frac{\pi v}{\nu}\int_{-\mathcal{L}_l/2}^{\mathcal{L}_l/2}\!\!\!\mathrm{d}x \,\rho^2_l(x),
\end{equation}
where $v$ is the edge magnetoplasmon velocity and ${\mathcal L}_l$ is the edge length.
The theory is bosonized with the prescription $\rho_l(x)=\partial_x\phi_l(x)/2\pi$ where $\phi_l(x)$ are scalar fields
comprising both a charged and a neutral sector:
\begin{equation}
\phi_l(x)=\phi_l^0(x)+\phi_l^p(x).
\end{equation}
The direction of motion of the fields is fixed by the external magnetic field. Here we
chose for convenience to set curvilinear abscissas in such a way that all fields have the same chirality (right movers).
The periodic neutral plasmonic mode  for the edge $l$  is
\begin{equation}\label{fipi}
\phi_l^p(x)=\sum_{{k_l}>0}\sqrt{\frac{2 \pi \nu}{k_l \mathcal{L}_l}}(a_{l,{k_l}}^{}e^{i k_l x}+a_{l,{k_l}}^\dagger e^{-i k_l x}) e^{-k_l a/2},
\end{equation}
where $a$ is an ultraviolet cutoff and bosonic creation and annihilation operators obey
$\lbrack a_{i,k}^{},a_{j,k'}^\dagger \rbrack = \delta_{ij}\delta_{kk'}$ with a
quantized wavevector $k_l=m \frac{2\pi}{\mathcal{L}_l}$, $m \in\mathbb{N}$.
The charged zero mode reads
\begin{equation}\label{zeromode}
\phi_l^0(x)=\frac{2 \pi }{\mathcal{L}_l} \nu n_l x - \chi_l
\end{equation}
with $n_l$ the excess number of quasiparticles  and $\chi_l$ an Hermitian operator conjugate to $n_l$.
The canonical commutation relation  $\lbrack \chi_j,n_l \rbrack = i \delta_{jl}$ together with
$\lbrack \phi_j^0(x),\phi_l^p(x')\rbrack=0$
ensure that the field  on each edge satisfies
\begin{equation}\label{commphi}
\lbrack \phi_l(x),\phi_{l}(x')\rbrack=i \pi \nu \sgn(x-x').
\end{equation}
Charge $\nu$ quasiparticle fields are now defined through exponentiation,
\begin{equation}
\label{quasip}
\psi_l(x)=\frac{1}{\sqrt{2\pi a}} e^{i  \phi_l(x)}e^{i\pi\nu x/\mathcal{L}_l  },
\end{equation}
where the extra phase has been added to preserve the correct twisted boundary conditions\cite{fradkin} for the qp field,
$\psi_l(x+\mathcal{L}_l)=
\psi_l(x)e^{i2\pi n_l \nu}$.
Equation (\ref{commphi}) guarantees that the fields $\psi_l$ create fractional charge excitations.
The same commutation relation also insures that quasiparticles have fractional statistics
\begin{equation}
\psi_l(x)\psi_l^\dagger(x')=
\psi_l^\dagger(x')\psi_l(x)e^{i \pi \nu \sgn(x-x')}.
\end{equation}
For quasiparticles of different edges, fractional statistics is introduced with
suitable commutation relations  $[ \chi_j,\chi_l ]= i \pi \nu \sgn(w_j-w_l)$ with $w_R=-w_L=1$ and $w_A=0$.

With prescriptions (\ref{fipi}) and (\ref{zeromode}), Eq.~(\ref{hamilt}) becomes
  \begin{equation}\label{energies}
 H^0_{{l}} =E_{\rm c}^l n_l^2 +\sum_{s=1}^\infty s \epsilon_l a_{l,s}^\dagger a^{}_{l,s},
 \end{equation}
 where $E_{\rm c}^l=\pi  \nu v/\mathcal{L}_l$ is the topological charge excitation energy;
 for the neutral sector,  $\epsilon_l= 2\pi  v/\,\mathcal{L}_l$ is the plasmonic excitation energy.

The total excess charge on an edge is given by
\begin{equation}\label{charge}
Q_l=e \int_{-\mathcal{L}_l/2}^{\mathcal{L}_l/2}\!\!\!\mathrm{d}x\,  \frac{\partial_x \phi_l(x)}{2\pi}=n_l e^*,
\end{equation}
and is conserved in the absence of tunneling.

Finally, the limit  $\mathcal{L}_L, \mathcal{L}_R \to\infty$ is taken.
In the following, for brevity's sake,
we relabel the antidot variables as  $\mathcal{L}=\mathcal{L}_A$, $\epsilon=\epsilon_A$,  $E_{\rm c}=E_{\rm c}^A$ and $n=n_A$.
\subsection{Aharonov-Bohm coupling}
The Hamiltonians Eq.~(\ref{hamilt}) are expected to describe the system
anywhere on a $\nu=1/p$ plateau since they are based on incompressibility.
Nevertheless the antidot edge, encircling a finite area, is sensitive to the actual position in the plateau through a coupling
to the Aharonov-Bohm (AB) flux.\cite{GellerLoss2,gold} We model this effect with an extra magnetic field
pointing in the opposite direction with respect to the background magnetic field.\cite{nota3}
The AB vector potential along the antidot edge reads $\vert \vec{A}\vert= \frac{\Phi}{\mathcal{L}}$
where $\Phi$ is the AB flux of the additional magnetic field.
The Aharonov-Bohm coupling is
$H^{\rm AB}\propto \vec{j}\cdot \vec{A}$ and describes the coupling of the antidot current density
with $\vec{A}$. By a gauge transformation, it is easy to see\cite{Geller97} that this amounts simply to a shift in the energies in
$H^0_{{A}}$ in Eq.~(\ref{energies})
 according to $E_{\rm c} n^2\to E_{\rm c} (n-\Phi/\Phi_0)^2$, where
  $\Phi_0=hc/\vert e\vert$ is the flux quantum.
  \subsection{Tunnel coupling}\label{tunnelcoupling}
Each $\chi$LL supports several excitations other than the single quasiparticle Eq.~(\ref{quasip}),
given in general by $\psi_l^m(x)\propto \exp[i m \phi_l(x)], \, m\in \mathbb{N}$. The electron corresponds
to $m=1/\nu$, while the single quasiparticle fields $\psi_j(x)$ are obtained by setting $m=1$.
One should therefore consider all possibilities for tunneling,
i.e. all terms like $ t^{(m)}{\psi_A^m}^\dagger(x) \psi_{j}^m(x)$, $j=L,R$.

Renormalization group (RG) flow equations have been set up for the
antidot geometry in Ref.~\onlinecite{GellerLoss2}. Here
we remind that the renormalized $m$-quasiparticle tunneling amplitude $t^{(m)}_{\rm ren}$ scales as
a power law
\begin{equation}\label{ren}
t^{(m)}_{\rm ren}=\left( \frac{a_{\rm ren}}{a}\right)^{1-m^2 g} t^{(m)}
\end{equation}
under increase in the unit-cell size $a \to a_{\rm ren}$.
Considering e.g. $g=1/3$, the single-qp tunneling amplitude $t^{\rm qp}=t^{(1)}_{\rm ren}$ diverges
when scaling to lower energies, while the electron amplitude $t^e=t^{(3)}_{\rm ren}$  scales to zero.
The largest $a_{\rm ren}$ attainable is the minimum between the thermal length $\propto v/T$
and the antidot length $\mathcal{L}$. 
A crossover temperature  $k_{\rm B}T_0\sim E_{\rm c}=\pi \nu v/\mathcal{L}$  exists such   that
for $T>T_0$ one has \mbox{$t^{(m)}_{\rm ren}/t^{(m)} \propto T^{m^2 g -1}$}, while for $T<T_0$ the flow is cutoff by
the energy associated to the finite size of the antidot $t^{(m)}_{\rm ren}/t^{(m)} \propto E_{\rm c}^{m^2 g -1}$.
In the following we will assume  $k_{\rm B}T \ll E_{\rm c}$ and bare tunneling amplitudes such that
electron tunneling can be neglected. Indeed this appears to be the case in most experimental observations where
single quasiparticle tunneling is clearly observed.\cite{gold}
We  therefore only retain the dominant term
\begin{equation}
H^T=\sum_{j=L,R} H^{{T}}_j= v \sum_{j=L,R}
\left(t_{{j}}\psi_{{A}}^\dagger(x_{j})\psi_{{j}}(0)  + h.c.\right),
\end{equation}
that represents the  single-quasiparticle tunneling between the infinite
edges and the antidot. Here the velocity $v$ is introduced to have dimensionless tunneling amplitudes $t_j$.

A finite source-drain voltage $V$ is applied between the left and right edges, producing a backscattered tunneling
current $I(t)$ of quasiparticles through the antidot
\begin{equation}\label{current}
I(t)=\left[\dot{Q}_L(t)-\dot{Q}_R(t)\right]/2,
\end{equation} with $\dot{Q}_{j}(t)=i[Q_j,H]= -ie^* v \lbrack   t_{j} \psi_{{A}}^\dagger(x_{j},t)\psi_{{j}}(0,t) - h.c. \rbrack$
and $j=L,R$.
We will consider asymmetric voltage drop  with $\alpha V$ and $(1-\alpha)V$ the voltage drops across the
left and right barrier and $0<\alpha<1$ (see Fig.~\ref{figuretwo}a)).
\section{Method}
\subsection{Sequential tunneling rates}\label{sec:tunn}
\begin{figure}
 \includegraphics[angle=0,width=8.5 cm]{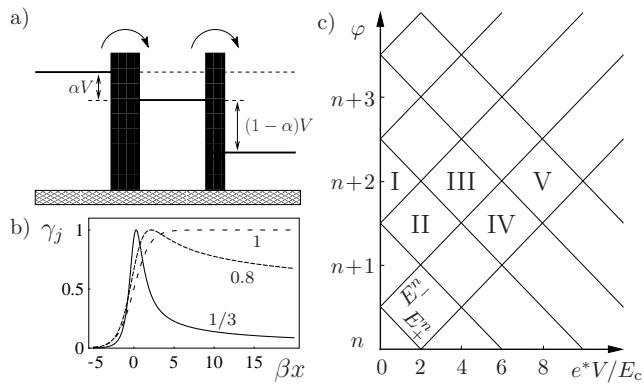}
 \caption{a)  Double-barrier system, with the voltage drop $\alpha V$ [$(1-\alpha)V$)]
    on the left [right] barrier. The forward transitions,
    corresponding to qp transfers contributing positively to the current,
    are indicated.  b) Plot of
    $\gamma_j(x)$ in Eq.~(\ref{ratedot}) as a function of $\beta x$ for $g=1/3,0.8,1$, in units
   $\vert t_j\vert^2 (\omega_c/4\pi^2)(\beta\omega_c/2\pi)^{1-g}/\mathbf{\Gamma}(g)$
and normalized to its maximum value.
   c) Scheme of transport regions in the $(V,\varphi)$ plane. Roman numbers
  indicate  the number of charge states involved in the transport.
  Thin lines signal the onset of transitions, where energies $E_\pm^n=0$(see Eq.~(\ref{enpiumeno})).
  }
 \label{figuretwo}
\end{figure}
For small tunneling as compared with temperature, transport can
be safely described within the sequential tunneling regime.\cite{furusaki}
Here, the main ingredients are the incoherent tunneling rates $\Gamma_{L,R}(E)$.
They are obtained from the transition probability
between the antidot state with $n$ qp at time $0$ and the state with
$n'$ qp at time $t$, at second order in the tunneling Hamiltonian\cite{braggioEPL}
\begin{equation}\label{rates}
\Gamma_j(E)= \left(\frac{v}{2\pi a}\right)^2\vert t_j\vert^2\int_{-\infty}^{+\infty}
\mathrm{d}\tau  e^{-W_A(\tau)}e^{-W_j(\tau)}e^{i\tau E},
\end{equation}
where
\begin{equation}
 W_{l}(t)=-\langle \phi_{l}^p(x,t)\phi_{l}^p(x,0) \rangle + \langle \phi_{l}^p(x,0)\phi_{l}^p(x,0) \rangle
\end{equation} are the
thermal correlation functions for the edge  ($L,R$)  and antidot ($A$) plasmonic excitations.
They can be cast in the standard dissipative form\cite{braggioEPL}
\begin{equation}
W_l(t)=\int_0^\infty \!\!\!\mathrm{d}\omega \frac{J_l(\omega)}{\omega^2}\left(\left(\cos \omega t-1\right)\coth\frac{\beta\omega}{2}-
i \sin \omega t\right)
\end{equation}
with the antidot and lead spectral densities
\begin{subequations}
\begin{eqnarray}
J_A(\omega) &=&\tilde{g} \omega \epsilon \sum_{s=1}^{\infty}\delta(\omega-s\epsilon) e^{-\omega/\omega_c}\label{densantidot}\\
J_{L,R}(\omega)&=&g \omega e^{-\omega/\omega_c}\label{densleads}
\end{eqnarray}
\end{subequations}
where $\omega_{c}=v/a$ is a high energy cut-off and
$\beta=1/k_{\rm B}T$. It is important to observe that this expression holds with the assumption that the
plasmonic fields $\phi_l^p$ are fully relaxed to thermal equilibrium. Mechanisms that could guarantee this
assumption include for instance thermalizing interactions with external degrees of freedom.
In the standard $\chi$LL
theory $g=\tilde{g}=\nu$ for all edge fields. Here we consider
the possibility that $g,\tilde{g}\neq\nu$ to describe renormalization effects.

In the present paper, we do not enter into a microscopical
derivation of exponents and only assume phenomenologically that
renormalization of exponents takes place. The functional form of the
quasiparticle correlators is thus preserved but is governed by a
parameter $g=\nu F$. The explicit value of $F$ will depend on the
details of interaction,  and we will consider it as a parameter.

Several mechanisms have been proposed to account for renormalization.
A striped phase is analysed in Ref.~\onlinecite{halpros}, where interaction is assumed between a $\nu=1/p$ $\chi$LL and phonon-like
excitations carrying no net electric current.
No tunneling of charge takes place between the $\chi$LL and the extra phonon modes, nor between the latter and the contacts. This
assumptions are relevant in the case of tunneling between two edge states across a Hall liquid, as in the weak backscattering limit
of the point-contact geometry or of our edge-antidot-edge setup. The only role of the additional modes
is a modification of the backscattering dynamics between two
chiral edges, leading to a renormalized exponent $g=\nu F$, where $F>1$ is a function of the coupling strength and of the phonon sound velocity.

Another possibility for tunneling renormalization relies on density-density interactions
between chiral edge states in a split Hall bar geometry, with
parameters $g_1$ and $g_2$ describing respectively  coupling across
the constriction and across the Hall bar.\cite{papa} Correlation functions for
quasiparticles can be calculated exactly in this model and give rise
to an interedge tunneling current governed by $g=\nu F$,
where now $F$ depends on $g_1$ and $g_2$.

Other proposals have been put forward and point toward more
profound modifications of standard chiral Luttinger Liquids. Interactions with phonons have been considered,
causing the bosonic field $\phi$ to split into several normal modes.\cite{eggert,khlebnikov}
Edge reconstruction offers a further scenario, with a non-monotonic density profile of the two-dimensional Hall droplet near
the edge essentially induced by electrostatics.\cite{evers,goldmant,joglekar,tsiperg,wan,yang}

We now return to the expression of the rates Eq.~(\ref{rates}). It is well known within the
bosonized description of edge
states and reads\cite{GellerLoss2,braggio,braggioEPL,matson}
\begin{equation}\label{rate}
\Gamma_j(E)= \sum_{s=-\infty}^{+\infty} w_s \gamma_j(E-s\epsilon),
\end{equation}
where
\begin{equation}\label{ratedot}
\gamma_j(x)= \vert t_j\vert^2 \frac{\omega_c}{(2\pi)^2}
\left( \frac{\beta\omega_{\rm c}}{2 \pi}\right)^{1-g}  \frac{\vert\mathbf{\Gamma}(g/2+i\beta x/2\pi)\vert^2}
{\mathbf{\Gamma}(g)} e^{\beta x/2} 
\end{equation}
with $\mathbf{\Gamma}(x)$ the Euler Gamma function.
 The sum in Eq.~(\ref{rate}) represents the contribution of
 plasmons, with  weight factors $w_s$.\cite{braggioEPL} At $T=0$ they are
\begin{eqnarray}\label{weights}
w_s&=&\frac{\mathbf{\Gamma}(\tilde{g}+s)}{\mathbf{\Gamma}(\tilde{g})s!}
\left(\frac{\epsilon}{\omega_c}\right)^{\tilde{g}}
e^{-s\epsilon/\omega_c} \Theta(s),
\end{eqnarray}
with $\Theta(s)$ the Heaviside step function.

It is apparent that the rates have contributions from
both the edge and the antidot correlators.
Their functional behaviour  is
greatly influenced by the value of the \emph{lead} Luttinger parameter $g$.
 In particular, for $g<1$ they present a non-monotonic behaviour
as a function of energy (Fig.~\ref{figuretwo}b)).

Note also that the fractional charge $e^*$ is solely determined by $\nu$ and is thus separated from the dynamical behaviour
governed by $g$: it is this separation that allows to find independent signatures of $\nu$ and $g$.

We now specify the temperature regime where our sequential tunneling picture holds.\cite{furusaki}
A higher limit is set by the condition $k_{\rm B}T\ll E_{\rm c}$, necessary to have a well-defined number of qp in the antidot
against thermal fluctuations. We then observe that the rates scale at low temperatures as $T^{g-1}$
at energy $E=0$. Hence, the linear conductance maximum is
$G_{\rm max}=C G_0 (\beta\omega_c)^{2-g}$,
with  $G_0=e^2/h$ and $C$ a constant parameter.\cite{braggioEPL}
For $g<1$, $G_{\rm max}$
increases with decreasing temperature, implying that for extremely low temperatures
transport is better described in the opposite regime of weak electron tunneling.\cite{kanefisher92} Sequential tunneling
approximation holds if $G_{\rm max}\ll G_0$. This implies $T
\gg T_{\rm min}$, with $T_{\rm min}$ that can be
extracted from a knowledge of typical measurements of $G_{\rm
max}/G_0$ at fixed temperature. For instance, from
experiments\cite{gold} performed at $T\approx 10$ mK with $G_{\rm
max}/G_0\approx 10^{-2}$ and $\nu=1/3$, assuming $g=\tilde{g}=\nu$ we estimate $T_{\rm min}\approx 1$ mK and
$E_{\rm c}\approx 120$ mK, so that the range between lower and higher
limits $T_{\rm min}\ll T\ll E_{\rm c}/k_{\rm B}$ spans two full decades in temperature.
The validity range can
be extended for systems with smaller antidot size (increased $E_{\rm c}$)
and weaker edge-antidot couplings (decreased $C$).
\subsection{Moments}
Hereafter, we will introduce higher current moments
as a tool to determine the $\chi$LL exponent and the carrier charge.
In particular, we will consider the current and the $p$-th normalized current cumulant,\cite{LR01} for $p=2,3$
 \begin{eqnarray}
\label{norm-moments}
k_p=\frac{\langle  I  \rangle_p }{|e^{p-1} \langle I  \rangle |} .
\end{eqnarray}
Here, $\langle I \rangle_p$
is the $p$-th irreducible current moment  and $\langle I  \rangle$ is the stationary current.
They can be expressed\cite{LR01}
in terms of the irreducible moments of
the  number $N$ of  charges $e^*$  transmitted in the  time $\tau$: $\langle I \rangle_p=\lim_{\tau\to\infty}\vert e^*\vert^p
\langle N \rangle_p /\tau$.
Observing that $\langle I  \rangle=\langle I  \rangle_1$
one has
\begin{equation}\label{momentofqp}
k_p= \left( \frac{e^*}{e} \right)^{p-1}   \frac{ \langle N   \rangle_p } { \langle N\rangle_1 }.
\end{equation}
Fano factor and normalized skewness correspond to $k_{2,3}$ respectively.
They are expressed as a product of two contributions:
 one coming from the charge of the carrier and the other from the statistics of the transport process, given in terms of \emph{particle} number
irreducible moments.

Our task will be to find conditions where $\langle N  \rangle_{p} / \langle N \rangle_1$  assumes \emph{universal}
values, independently from e.g.  the tunneling amplitudes $t_j$ and the asymmetry $\eta=\vert {t}_R\vert^2/\vert {t}_L\vert^2$.
We put particular emphasis on the fact that such universality must hold independently of the Luttinger parameter $g$, since
renormalization processes are almost invariably present and not controllable.
 We define therefore \emph{special} the conditions in the parameter space where
 $\langle  N \rangle_p /  \langle N \rangle_1$ take universal values.
  Note that the statistics of a transport process is identified  by all its cumulants, and therefore
   all of them  should be required to be universal to identify \emph{special} regimes. Here, we will adopt
  only the minimal comparison of the second and third moment that are more accessible in experiments.
\subsection{Master equation}
The detailed analysis of  DC current and $k_{2,3}$ is obtained  from the cumulant
generating function calculated in the markovian master equation framework\cite{nazbag} in the sequential regime.
The occupation probability of a fixed number $n$ of antidot quasiparticles is\cite{GellerLoss2}
\begin{equation}\label{master}
\frac{\mathrm{d}p_n(t)}{\mathrm{d}t}=
\sum_{n'}\sum_j \left(   \Gamma_j(E_j^{n'\to n})p_{n'}(t) -\Gamma_j(E_j^{n\to n'}) p_{n}(t)\right). \end{equation}
The energies in the tunneling rates are the
differences between the antidot and edge $j$ energies before and after the tunneling event,
\begin{eqnarray}\label{enpiumeno}
E_L^{n\to n+1}&\!=\!&e^*V/2+ 2 E_{\rm c}(\varphi-n-1/2)\equiv E_+^n\nonumber\\
E_R^{n+1\to n}&\!=\!&e^*V/2- 2 E_{\rm c}(\varphi-n-1/2)\equiv E_-^n,
\end{eqnarray}
with $\varphi=\Phi/\Phi_0+(\alpha-1/2)e^* V/2E_{\rm c}$.
These forward transitions $n\to n+1$, $n+1\to n$ take place on the left and right barrier respectively (see Fig.~\ref{figuretwo}a)).
The corresponding backward energies obey
\begin{eqnarray}
E_L^{n+1\to n}&=& -E_L^{n\to n+1}\nonumber\\
E_R^{n \to n+1} &=& -E_R^{n+1 \to n}.
\end{eqnarray}
With $\Gamma(E)=\Gamma_j(E)/\vert t_j \vert^2$,
we define $\Gamma^n_\pm=\Gamma\left(E^n_\pm\right)$ and the corresponding
backward rate  $\overline{\Gamma}_\pm^n= \Gamma\left(- E^n_{\pm}\right)$.
Detailed balance states that $\overline{\Gamma}_\pm^n=e^{-\beta E^n_{\pm}}\Gamma_\pm^n$.

The conditions $E_\pm^n=0$ grid the $(V,\varphi)$ plane into diamonds according to the scheme in Fig.~\ref{figuretwo}c).
For the sake of clarity, in the following we consider a symmetric voltage drop
$\alpha=1/2$ when $\varphi$ depends on the dimensionless magnetic flux $\Phi/\Phi_0$ only.
 However, all results are cast in  generally valid form.
\begin{figure}
\includegraphics*[bb= 144 256 460 544,angle=0,width=8.5 cm]{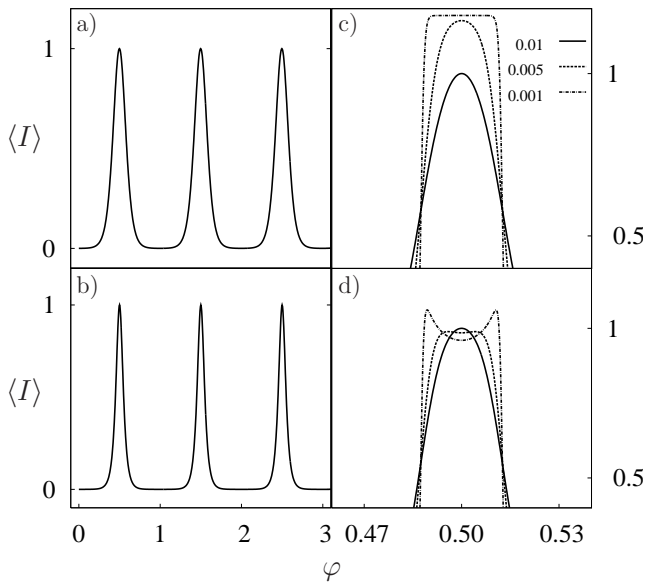}
\caption{Plot of the average current as a function of the magnetic flux
$\varphi$  for a) $g=\nu=1$, b) $g=\nu=1/3$ at  $k_{\rm B}T = 0.1 E_{\rm c}$.
  A zoom on the shape of the resonance peak is plotted for c) $g=\nu=1$ and
 d) $g=\nu=1/3$ with different temperatures
$k_{\rm B}T = 0.01,0.005,0.001  E_{\rm c}$.
Other parameters: $e^*V = 0.05 E_{\rm c}$, $\eta=1$.
Curves in panels a) and b) are normalized to the
on-resonance value, while curves in panels c) and d) are normalized
to the maximum value of the current at $k_{\rm B}T=0.01E_{\rm c}$ (solid
line).} \label{figurethree}
\end{figure}
\section{Results}
This Section is organized as follows: we describe current, Fano factor and skewness assuming $g=\tilde g$,
and we discuss various limits in the temperature, voltage and flux range,
referring to the diamonds in Fig.~\ref{figuretwo}c).
\subsection{Current}
A great deal of information on the interaction parameter can be gathered from the analysis of the DC current, although it is not
possible to extract signatures of the fractional charge.
In this subsection we consider $g=\nu$, despite the generalization $g\neq\nu$ is straightforward.
\subsubsection{Magnetic flux dependence}
In the low-voltage regime $e^*V\lesssim 2E_{\rm c}$ (diamonds I and II in Fig.~\ref{figuretwo}c)) a simple analytical
form for the current can be found; here, only two adjacent antidot charge states  are connected through open rates and
charge transport takes place in a sequence of $n \to n+1\to n$ processes.

In regions I transport is exponentially suppressed because of the finite energy $E_{\rm c}$;
in regions I,II current at fixed source-drain voltage oscillates as
 a function of $\varphi$  with a periodicity of one flux quantum $\Phi_0$
for \emph{any} $\nu$ and $g$, in accordance with gauge invariance.\cite{byeryang}
This is represented  in Fig.~\ref{figurethree}a) and b) obtained from a numerical solution of the master equation.

Because of periodicity in $\varphi$, we start at $n=0$.
The current is
\begin{equation}\label{corr}
\langle I \rangle= \eta \vert e^*\vert  \vert t_L \vert^2\frac{\Gamma_+^0 \Gamma_-^0 f_-(e^* V)}{\Gamma_{\rm tot}},
\end{equation}
where
\begin{eqnarray}\label{twostate}
\Gamma_{\rm tot}&=& \Gamma_+^0 f_+(E_+^0) + \eta \Gamma_-^0 f_+(E_-^0),\nonumber\\
f_\pm(x)&=&1\pm e^{-\beta x}.
\end{eqnarray}

If  temperature is lowered until $k_{\rm B}T\ll e^*V$, then a qualitative difference appears in the
shape of the resonance peaks as a function of $g$.
When $g<1$, due to the non-monotonic rates
the current develops two side peaks across the resonance and develops a minimum on resonance
(Fig.~\ref{figurethree}d)).
On the other hand, for $g=1$ the rates Eq.~(\ref{ratedot}) are Fermi functions and the
current peak displays no structure regardless of the ratio $e^*V/k_{\rm B}T$.
The only effect is that for lower $T$ plateaus appear with increasing width (Fig.~\ref{figurethree}c)).

It is worth mentioning that the double-peak structure for $g<1$ was
already established in resonant qp tunneling through localised
impurities,\cite{chamon} although in our case the parameter that
tunes the resonance is the magnetic flux.

\begin{figure}[h!]
\includegraphics[angle=0,width=8.5 cm]{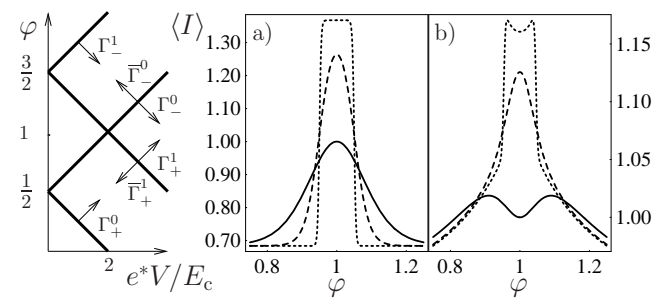}
 \caption{Left panel: scheme of the retained rates in the approximation for the three-state region. Right panel: a)
 current  versus $\varphi$ at $g=\nu=1$ for $\beta E_{\rm c}=10,25,200$, with solid, dashed, dotted  curves respectively.
 The curves are at fixed $e^*V/E_{\rm c}=2.2$ and $\eta=1$, and are normalized to the value
 of the lowest temperature curve at $\varphi=1$. b) the same with $g=\nu=1/3$.}
 \label{figurefour}
\end{figure}

If bias voltage is increased, one enters regimes where more charge states
participate to transport. For voltages $2 E_{\rm c}<e^*V< 4E_{\rm c}$ one needs to consider
two or three charge states. As an example of regions III, we discuss the diamond where
the rates $\Gamma^p_\pm,\overline{\Gamma}_\pm^p$, $p=n,n+1$,
are open to transport and the states with $n,n+1$ and $n+2$ qp are involved.
Again, periodicity allows to choose the starting point $n=0$.
Furthermore, in our temperature regime only two backward rates can be retained, $\overline{\Gamma}_-^0$ and
 $\overline{\Gamma}_+^1$  (Fig.~\ref{figurefour}, left panel). One finds:
\begin{equation}\label{currthree}
\langle I \rangle = \eta \vert e^*\vert  \vert t_L \vert^2 \frac{\Gamma_+^1\Gamma_-^1\Gamma_t^0 +\Gamma_+^0\Gamma_-^0\Gamma_t^1}{
\Gamma_t^0(\Gamma_+^1+\Gamma_t^1)+\eta \Gamma_-^0\Gamma_t^1},
\end{equation}
where $\Gamma^0_t=\Gamma_+^0 + \eta \overline{\Gamma}_-^0$ and
$\Gamma^1_t=\eta\Gamma_-^1 +   \overline{\Gamma}_+^1$.
Here also double-peak structures appear at fixed bias voltage as a function $\varphi$, although they have a more
complicated behaviour. Indeed, close to the diamond onset, $e^*V\approx 2E_{\rm c}$, the current for $g<1$
develops a broad minimum around $\varphi=1$ for large temperature, then a maximum in an intermediate range, and finally
a minimum again with sharp side-peaks at $\beta E_{\rm c}\gg 1$ (Fig.~\ref{figurefour}, right panel b)).
No structure appears for $g=1$, where decreasing the temperature
has the only effect of shrinking the peak width and causing a larger plateau (Fig.~\ref{figurefour}, right panel a)).
\subsubsection{Bias voltage dependence}
\begin{figure}[h!]
\includegraphics*[bb= 148 266 365 497,width=8 cm]{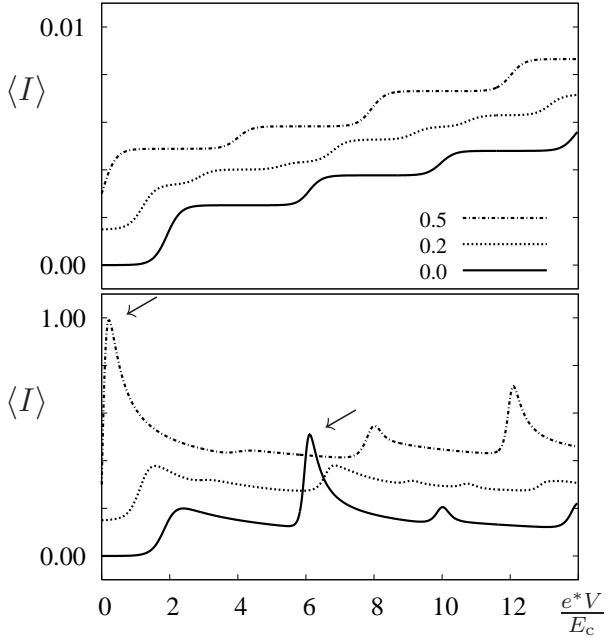}
 \caption{Current as a function of voltage  for $g=\nu=1$ (top) and $g=\nu=1/3$
(bottom). The lines correspond to $\varphi=0,0.2,0.5$  and are shifted of $0.0015 I_0$ ($0.15 I_0$) with respect
to the curve at $\varphi=0$. Other parameters are $\eta=1$
 and $k_{\rm B}T=0.1 E_{\rm c}$. Current units are $I_0= \vert e^*\vert
\omega_c \vert t_L \vert^2(\beta \omega_c)^{1-g}(2\pi)^{g-3}   /\mathbf{\Gamma}(g)$. }
 \label{figurefive}
\end{figure}

A plot of the current is presented in Fig.~\ref{figurefive} as a function of the source-drain voltage for symmetric
barriers. Again the rate behaviour for $g=1$ or $g<1$ changes qualitatively the current.
While for $g=1$ the current reflects the Fermi-liquid nature of the leads and increases in steps,
non-monotonic features appear for $g=1/3$ and lead to negative differential conductance (Fig.~\ref{figurefive} b)).
Two features are most remarkable in this sense: the on-resonance peak (curve at $\varphi=0.5$) for small voltage and the off-resonance peak
(curve at $\varphi=0$) at $e^*V\approx 6E_{\rm c}$, indicated with arrows in Fig.~\ref{figurefive} b).
The first peak develops in region II and can be described with Eq.~(\ref{corr}). Setting $\varphi=1/2$ and $\eta=1$, one finds
\begin{equation}
\langle I \rangle=\vert e^* \vert \vert t_L \vert^2 \frac{\Gamma^0_+}{2} f_-\left(\frac{e^*V}{2}\right).
\end{equation}
This result allows for a determination of the Luttinger parameter $g$ from the
power-law behaviour of the rate, $\Gamma^0_+ \propto \left(\beta e^* V /4\pi\right)^{g-1}$, valid for $\beta e^*V \gg 1$.

The second feature finds a natural explanation in terms of antidot
plasmonic excitations. These in fact enter into play for
$e^*V>\epsilon=2E_{\rm c}/\nu$. This new mode increases non-monotonically the tunneling rate, leading
to peaks in current.

\begin{figure}[h!]
\includegraphics*[bb= 105 209 547 530,angle=0,width=8  cm]{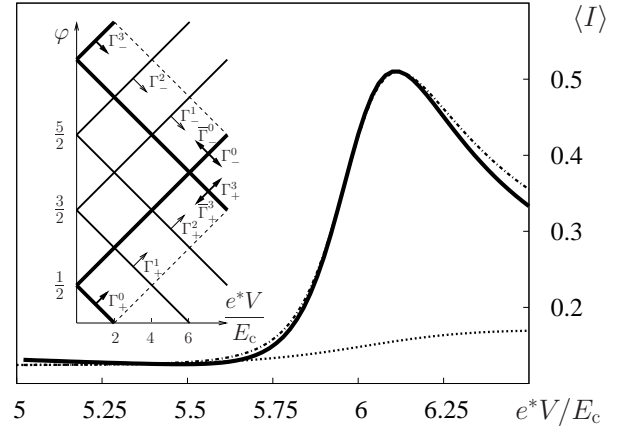}
 \caption{Current in the IV-V state regime versus bias voltage at $g=\nu=1/3$. Solid line, numerical
result; dot-dashed line, approximate solution in Eq.~(\ref{currapprox}) with $\delta\Gamma_p=g \Gamma^3_+$;
 dashed line, approximate solution without plasmons ($\delta\Gamma_p=0$).
 The curves are at $\varphi=2$ (equivalent to $\varphi=0$ in Fig.~\ref{figurefive} because of the periodicity in magnetic flux),
$k_{\rm B}T=0.1E_{\rm c}$, $\eta=1$ and units are as in Fig.~\ref{figurefive}. The best fit parameter is $f=0.19 I_0/\vert e^*\vert$.
 Inset: scheme of retained rates in the approximation for region V.}
 \label{figuresix}
\end{figure}

To ascribe the onset of the peak to plasmons,
we focus on the above example with $g=\nu=1/3$ and $\epsilon=2E_{\rm c}/\nu=6 E_{\rm c}$.
Here, one has to consider five charge states, $n,\ldots,n+4$ and
the rates $\Gamma^p_\pm$, $p=n, \ldots, n+3$.
In this regime it is possible to obtain an approximate form of the current
neglecting all backward rates except those opening at the border of region V ($\overline{\Gamma}^0_-,
\overline{\Gamma}^3_+$, see Fig.~\ref{figuresix}, inset).
We work at $\varphi=2$, choosing $n=0$ as a reference.
Observing that the large-energy behaviour of the forward rates is $\propto (\beta E)^{g-1}$,
we can process the rates according to the magnitude of their argument in region V: furthermore,
for $\eta=1$, $\Gamma_+^l=\Gamma_-^{3-l}$ with $l=0,1,2,3$.
Thus we set $\Gamma_\pm^1= \Gamma_\pm^2 \sim f$, with $f$ a
temperature dependent constant, since these rates are in their slow-decaying regime.
On the opposite, the full energy dependence of $\Gamma^3_+=\Gamma^0_-$ is
taken into account.
In terms of the 'plus' rates only, the current is given by
\begin{equation}\label{currapprox}
\langle I \rangle = 2\vert e^* \vert \frac{ f\, \overline{\Gamma}_+^3 +\Gamma^0_+ \left(\Gamma^3_+ + f \right)}
{3\Gamma^0_+ + 2\Gamma_+^3+3\overline{\Gamma}^3_+}.
\end{equation}
Finally, we set $\Gamma^0_+ \sim f + \delta\Gamma_{p}$, where $\delta\Gamma_{p}$
is the first plasmon contribution to the rate that shows up as a secondary peak at energy $E=\epsilon$.
We now compare the current with no plasmons ($\delta\Gamma_p=0$)
with $\delta\Gamma_p=g \Gamma^3_+$ obtained from Eqs.~(\ref{rate}) and (\ref{weights})
with relative weight $w_1/w_0=\mathbf{\Gamma}(g+1)/\mathbf{\Gamma}(g)=g$ .
Figure \ref{figuresix} shows a comparison between the current with these approximations and the
result obtained through a numerical solution of the master equation.
It is apparent that neglecting the plasmon completely misses the peak onset that is instead well-captured
by out latter approximation. This testifies
that plasmonic excitations play indeed the crucial role in current enhancement.
\subsection{Current moments}\label{currmom}
Current moments are now discussed.\cite{merlo}
According to Eq.~(\ref{momentofqp}), these two quantities are the natural observables to look at
in order to measure the quasiparticle charge $e^*$. In the antidot
geometry, different conditions can be found where this measurement is \emph{special} in the sense
discussed above. We remind that for a poissonian process, as with
weak backscattering current in a point contact, $k_2=e^*/e=\nu$, $k_3=(e^*/e)^2$.
\subsubsection{Few-state regime: $e^* V\lesssim 2E_{\rm c}$}
For the sequential tunneling master equation Eq.~(\ref{master})
restricted to the two-state regime (two charge states),
an exact analytical treatment is possible.

A known formula\cite{BlanterReview,braggio} for the Fano factor is obtained
\begin{equation}\label{fano}
k_2= \left(\frac{e^*}{e}\right)\left(\coth\left(\frac{\beta e^*V}{2}\right)-2 \eta \frac{ \Gamma_+^0 \Gamma_-^0 f_-(e^* V)}{\Gamma_{\rm tot}^2}\right),
\end{equation}
while for the skewness we find
\begin{eqnarray}\label{skewness}
k_3&=&\left(\frac{e^*}{e}\right)^2 \bigg(1-6 \eta \frac{ \Gamma_+^0 \Gamma_-^0 f_+(e^* V)}{\Gamma_{\rm tot}^2}+\nonumber\\
& & \qquad\qquad+ 12 \eta^2 \frac{  {\Gamma_+^0}^2 {\Gamma_-^0}^2 f_-^2(e^* V)}{ \Gamma_{\rm tot}^4}\bigg),
\end{eqnarray}
with the functions $\Gamma_{\rm tot},f_\pm$ defined in Eq.~(\ref{twostate}).
The previous equations are an example of the intertwining  between charge and process statistics that does not
allow, in general, for unequivocal conclusions on $e^*$.
We analyse now the behaviour varying the ratio $e^*V/k_{\rm B} T$ looking for \emph{special} conditions.\\
\emph{Thermal limit: $e^* V \ll k_{\rm B}T $.}
The Fano factor is independent of the charge fractionalization,
\begin{equation}
k_2=2 \frac{k_{\rm B} T}{e V},
\end{equation} reflecting the fluctuation-dissipation theorem.
On the contrary, the normalized skewness that measures the fluctuation asymmetry induced by the current
depends on the carrier charge $e^*=\nu e$.
Indeed, for  low voltages $V\to 0^+$  one has
\begin{equation}
\label{SkTerm}
k_3=\left(\frac{e^*}{e}\right)^2\left[1-3\frac{\eta}{(1+\eta)^2}\frac{1}{{\rm Cosh}^2( \beta E_{\rm c}(\varphi-1/2))}\right].
\end{equation}
Note that the $\varphi$ dependence can be used to extract $\eta$ and $e^*/e$ independently from $g$.\\
\emph{Shot limit: $k_{\rm B}T \ll e^* V$.}
 In the blockade regions I  with $|\beta E_\pm^0|\gg 1$,
 one has $k_{2}=\nu$ and $k_3=\nu^2$. In this
 case the statistics of the transport process is poissonian: the transport
 through the antidot is almost completely suppressed, $\langle I \rangle \approx 0$, and the residual
 current is generated only by a thermally activated tunneling that is completely uncorrelated. So
  regions I constitute an example of \emph{special} regime.
\begin{figure}
\begin{center}
\includegraphics*[bb=175 257 522 522 ,width=8.6 cm,angle=0]{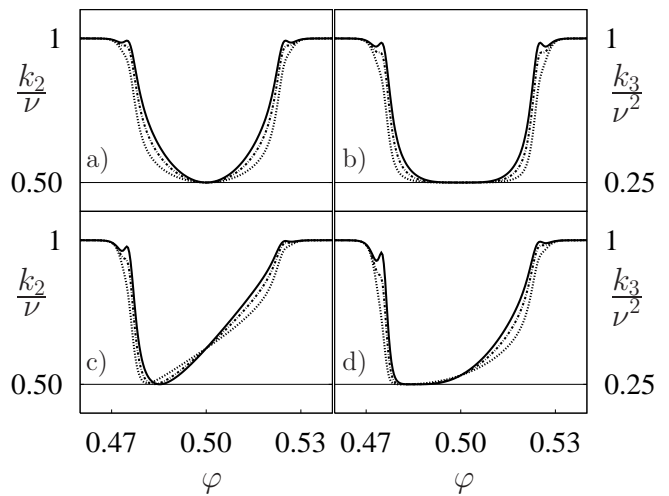}
\caption{ Fano factor (left) and skewness (right) as a function of $\varphi$. Panels a) and b)
$\eta=1$, panels c) and d) $\eta=3$. Parameters: $e^*V=0.1 E_{\rm c}$, $k_{\rm B}T=0.004 E_{\rm c}$ and
$g=1/5$ (solid line), $g=1/3$ (dot-dashed line), $g=1/2$ (dotted line). Horizontal thin lines indicate
the universal limits  at 1/2  at 1/4.}\label{figureseven}
\end{center}
\end{figure}
We consider now the two-state regime (II)  for $\beta E_\pm^0 \gg 1$.
For fractional edges $g<1$, $k_{2,3}$ have a particular functional dependence on $\varphi$.
We find that they both develop a minimum and that for not too strong asymmetries
 the absolute values of the minima are
 \begin{eqnarray}
 k_n^{\rm min}&=&\frac{\nu^{n-1}}{2^{n-1}}.
 \end{eqnarray}
 These minimal values do not depend neither on $g$,
as the comparison of solid ($g=1/5$),  dotted ($g=1/3$) and dashed ($g=1/2$) curves in Fig.~\ref{figureseven} confirms, nor on $\eta$.
For Fermi liquid edges $g=1$, we have
\begin{eqnarray}
k_2&=&\nu(1+\eta^2)/(1+\eta)^2 \nonumber\\
k_3&=&\nu^2\left[1-6\eta(1+\eta^2)/(1+\eta)^4\right],
\end{eqnarray}
independently from $\varphi$.
Here, $k_2$ and $k_3$  assume their minimal values $\nu/2$ and $\nu^2/4$  in the symmetric case $\eta=1$.
In this conditions we have the strongest anticorrelation that is signalled by a marked sub-poissonian statistics.

We can conclude that in the two-state regime, in the shot limit, the values of the minima for $k_{2,3}$
obtained varying $\eta,\varphi$ correspond to a \emph{special} condition where the system shows
the same universal sub-poissonian statistics  for any $g\leq1$.
This represents a means of testing fractional charge outside poissonian conditions and insensitive to renormalizations of the
Luttinger parameter.

In the intermediate regime $e^* V \approx k_{\rm B}T $, $k_{2,3}$ depend more strongly on the parameter $g$, and the interplay
of these two energy scales prevents the onset of special regimes.
\subsubsection{Many-state regime: $e^*V> 2E_{\rm c}$}
\begin{figure}
\begin{center}
\includegraphics[width=8.6 cm]{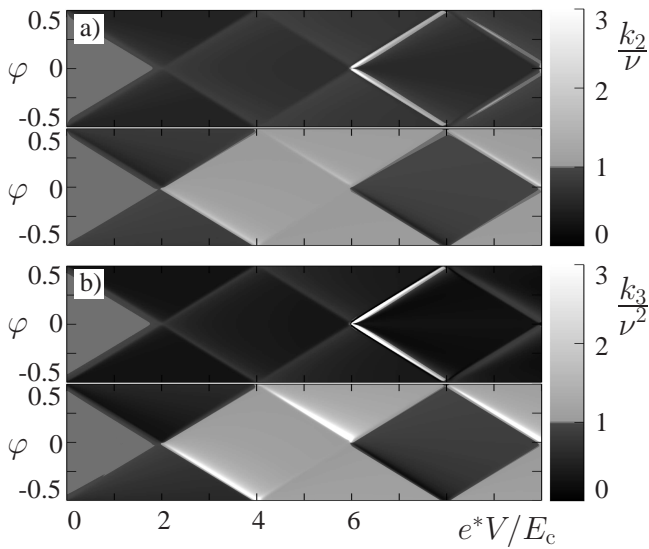}
\caption{Panel a): Fano factor $k_2/\nu$ at $\nu=g=1/3$, $k_{\rm B}T=0.02 E_{\rm c}$, vs. source-drain voltage and $\varphi$.
Panel b): normalized skewness $k_3/\nu^2$ with the same parameters.
Both moments are plotted
for symmetric barriers $\eta=1$ (top) and strong asymmetry $\eta=10$ (bottom).
}\label{figureeight}
\end{center}
\end{figure}

We study now higher voltages $e^*V > 2E_{\rm c}$
where the renormalized interaction parameter $g$ has a prominent role.
For this purpose we consider the behaviour of both the Fano factor and the normalized skewness.
In Fig.~\ref{figureeight} a density plot of  $k_2$ and $k_3$ for $\nu=g=1/3$
as a function of magnetic flux and source-drain voltage is shown for different asymmetries.
First of all, in region I we recover  poissonian statistics: $k_2=\nu$, $k_3=\nu^2$  (middle grey) for any $\eta$.
 Outside this region, light grey zones represent super-poissonian
Fano factor and  normalized skewness ($k_2>\nu$ and $k_3>\nu^2$), while dark grey regions
represent sub-poissonian behaviour ($k_2<\nu$ and $k_3<\nu^2$).
Figure \ref{figureeight} shows that super-poissonian regions are possible with
$g<1$. For $g\geq1$, we always have sub-poissonian behaviour in accordance
with previous results (not shown).\cite{braggio}
In presence of asymmetry, super-poissonian regions increase. We note that Fano and skewness present
concurrent super/sub-poissonian behaviour, and
that the maximal values of the skewness in the super-poissonian regions are stronger.
Given this similarity, in the following we will discuss the Fano factor only. \\
\emph{Three-state region}. In region III, a
tractable analytical formula for the Fano factor can be derived under the same assumptions made for the current Eq.~(\ref{currthree}):
   one has $k_2/\nu=1-2\eta\,\,\delta k_2 $, with
\begin{equation}\label{deltak}
 \delta k_2  \!=\! \frac{  {\Gamma^0_t}^2\Gamma_+^1 \Gamma_-^1 \!+\!{\Gamma^1_t}^2\Gamma_+^0 \Gamma_-^0 \!+\!
 \Gamma_-^0 \Gamma_+^1 \left(\Gamma^0_t\!-\!\Gamma^1_t\right) \left(\eta\Gamma_-^1\!-\!\Gamma_+^0\right) }
  {\left[ \eta \Mvariable{ \Gamma_-^0}
         \Mvariable{ \Gamma_t^1} + \Mvariable{ \Gamma_t^0} \left( \Mvariable{ \Gamma_+^1} + \Mvariable{ \Gamma_t^1} \right)  \right]^2},
\end{equation}
 with $\Gamma^0_t,\Gamma^1_t$  in Eq.~(\ref{currthree}).
 We note that in order to have super-poissonian noise a fractional  $g<1$ is necessary,
 with additional conditions on the asymmetry. Indeed, setting
 $\eta=1$ in Eq.~(\ref{deltak}) in the  limit $\beta E_+^1,   \beta E_-^0 \gg 1$
 yields $\delta k_2>0$ for any $g$.
 On the other side, setting $g=1$ gives $\delta k_2=2\eta/(\eta^2+\eta+1)^2>0$.
 It appears that positive correlations are induced by an interplay of $\eta$ and $g$.\\
\emph{Five-state region.} Finally, interesting effects take place in the five-state regime (V) for $\nu=1/3$.
Here, a  strongly super-poissonian Fano factor appears along the diamond lines
for $\eta=1$  and disappears for large asymmetries. An investigation of this effect can be performed with the
same methods used to obtain Fig.~\ref{figuresix}. The rate enhancement due to the onset of the plasmonic
collective excitations can again be shown to be responsible of the super-poissonian behaviour at small asymmetries.
\section{Conclusions}
In conclusion, we have analysed transport of quasiparticles
through an antidot coupled with edge states
in the fractional Hall regime. The model of a finite size chiral Luttinger liquid
with periodic boundary conditions has been reviewed and cast in a suitable form
for calculations of higher current moments through a master equation approach in the sequential
regime. We have also allowed for the possibility of a phenomenological
renormalization of the interaction parameter.

We have found that independent information on
interaction renormalization and fractional charge can
be extracted from tunneling current and its moments, noise and skewness.
For current, remarkable qualitative differences as a function of the Luttinger parameter
appear in the shot limit of resonance peaks and in the three-state regime. A quantitative
determination of $g$ is furthermore possible through the power-law behaviour of
on-resonance current versus voltage at low temperatures.
Current moments also depend strongly on the interaction parameter: in particular, super-poissonian
behaviour is never found for $g\geq1$ regardless of asymmetry.
On the other hand, we have identified \emph{special} regimes in the one- and two-state regions
where a comparison of Fano factor and normalized skewness realizes an unambiguous
charge determination procedure, insensitive to renormalizations of $g$.
Finally, signatures of plasmonic excitations are indicated in the large bias voltage regime.

Confirmation of such novel results appears to
be within reach: plugging estimated parameters from present
experiments\cite{gold} into our results gives currents in the range
of $0.5\div 5$ pA for $g=1/3$ and $T=10$ mK.
At the same time, recent accomplishments in measurement techniques applied to electron
counting open the possibility for feasible noise and skewness determination,
even in systems with very low current and noise levels.\cite{skewness}

Furthermore, litographic approaches have been developed that allow for antidot radii sensibly
smaller than the 300 nm in the original experiments.\cite{goldpriv}
Thus, the energy scale $E_{\rm c}$ associated with the antidot finite size  can be raised  to the order
of some hundreds of mK, and more easily attainable temperature regimes can be compatible with the requirements of our approximations.

We believe that our results help clarify some issues related to transport in the Hall regime,
namely features like fractional charges and nonpoissonian correlations. Such a thorough analysis of
a Hall antidot device is a necessary building block, especially in view of possible applications
of similar systems to topological quantum computation.\cite{averin}

We thank E. Fradkin, M. Heiblum, V. Pellegrini, J. Jain, V. Goldman, and M. Governale for useful discussions.
 We acknowledge partial support from ESF activity INSTANS.
Financial support by the EU via Contract No. MCRTN-CT2003-504574
and by the Italian  MIUR via PRIN05 is gratefully acknowledged.


\begin{thebibliography}{99}
\bibitem{nota3} This does not mean that a localized flux is actually needed in experiments. Such a modelization
simply takes into account the fact that the infinite edges are insensitive to the strength of the external magnetic
field as long as incompressibility is present. Indeed, in real experiments the antidot flux is changed sweeping the
external magnetic field, as in Ref.\onlinecite{gold}.
\end{thebibliography}
\end{document}